\input{aipcheck}

\documentclass[
    ,final            
  ]
  {aipproc}

\layoutstyle{6x9}

\begin{document}

\title{QCD Factorization for heavy quarkonium production 
at collider energies}


\author{Jian-Wei Qiu}{
  address={Department of Physics and Astronomy, Iowa State University\\
           Ames, Iowa 50011, U.S.A.\\
           Physics Department, Brookhaven National Laboratory\\
           Upton, New York 11973-5000, U.S.A.}
}

\begin{abstract}
In this talk, I briefly review several models of the heavy quarkonium 
production at collider energies, and discuss the status of QCD
factorization for these production models. 
\end{abstract}

\maketitle

\section{Introduction}
\label{intro}

The  production of bound states of heavy quark pairs 
has been the subject of a vast theoretical literature 
and of intensive experimental study, and 
offers unique perspectives 
into the formation of QCD bound states \cite{qwg-review}.
The first step in quarkonium production, the inclusive 
creation of a pair of heavy quarks, is an essentially 
perturbative process, and 
takes place at a distance scale
much smaller than the physical size of a quarkonium.  
Therefore, the transition from the produced heavy quark pairs 
to bound mesons is unlikely to be instantaneous.  
The ``long'' lifetime of the produced pairs allows 
the dynamics of QCD confinement to evolve, and 
provides us with a window of opportunities to probe the formation 
of QCD bound states via the interaction between the pairs
and the medium where they were produced.

Much of the predictive content of QCD perturbation theory
is contained in factorization theorems \cite{css-fac}.  
They allow us to separate long-distance physics from 
short-distance interactions
in hadronic cross sections, and to provide physical content 
for the uncalculable long-distance quantities, so that they 
can be measured independently or calculated numerically.
In this talk, I first briefly review several models of heavy 
quarkonium production, and then, discuss the status of QCD 
factorization for these production models, and finally, give
a brief summary and outlook.

\section{Production models}
\label{production}

In order to produce a heavy quarkonium in hadronic collisions, 
the energy exchange in the collisions has to be larger
than the invariant mass of the produced quark pair 
($\ge 2m_Q$ with heavy quark mass $m_Q$).  
The pairs should be produced at a distance scale
$\Delta r \le 1/2m_Q \le 0.1$~fm for charmonia or 0.025~fm for 
bottomonia.  
Since the binding energy of a heavy quarkonium of 
mass $M$ is much less than heavy quark mass, 
$(M^2-4m_Q^2)/4m_Q^2\ll 1$, the 
transition from the pair to a meson is sensitive to 
soft physics.  The quantum interference between the production
of the heavy quark pairs and the transition process 
is powerly suppressed by the heavy quark mass, and 
the production rate for a heavy quarkonium state, $H$, 
up to corrections in powers of $1/m_{Q}$, 
can be factorized as, 
\begin{equation}
\sigma_{A+B\rightarrow H+X} 
\approx 
\sum_{n} \int d\Gamma_{Q\bar{Q}}\
\sigma_{A+B\rightarrow Q\bar{Q}[n]+X}(\Gamma_{Q\bar{Q}}, m_Q) \
F_{Q\bar{Q}[n]\rightarrow H}(\Gamma_{Q\bar{Q}}) 
\label{qq-fac}
\end{equation}
with a sum over possible $Q\bar{Q}[n]$ states and an integration 
over available $Q\bar{Q}$ phase space $d\Gamma_{Q\bar{Q}}$.
Neglecting the ``high twist'' interaction between incoming hadrons
and quarkonium formation, the nonperturbative transition probability 
$F$ for a pair of off-shell heavy quark ($\psi$) and antiquark ($\chi$)
to a quarkonium state $H$ is proportional to the Fourier transform 
of following matrix elements
\begin{equation}
\sum_N
\langle 0|\chi^\dagger(y_1)\, {\mathcal K}_n\, \psi(y_2)|H+N\rangle
\langle H+N| \psi^\dagger(\tilde{y}_2)\, {\mathcal K}'_n\,
             \chi(\tilde{y}_1)|0\rangle\, ,
\label{F-nonlocal}
\end{equation}
where $y_i (\tilde{y}_i)$ are coordinates, and
${\mathcal K}_{n}$ and ${\mathcal K}'_{n}$ are 
local combinations of color and spin matrices for the $Q\bar{Q}$  
state $n$.  A proper insertion of Wilson lines to make
the operators in Eq.~(\ref{F-nonlocal}) gauge invariant is 
implicit \cite{nqs}.  
In Eq.~(\ref{F-nonlocal}),
$\sigma_{A+B\rightarrow Q\bar{Q}[n]+X}$  represents the production 
of a pair of on-shell heavy quarks and is calculable in perturbative 
QCD \cite{heavyquark}.  
The debate on the production mechanism 
has been focusing on the transition from the pair to the meson.

\begin{figure}
\begin{minipage}[c]{2.7in}
\begin{center}
  \includegraphics[width=2.4in,height=1.8in]{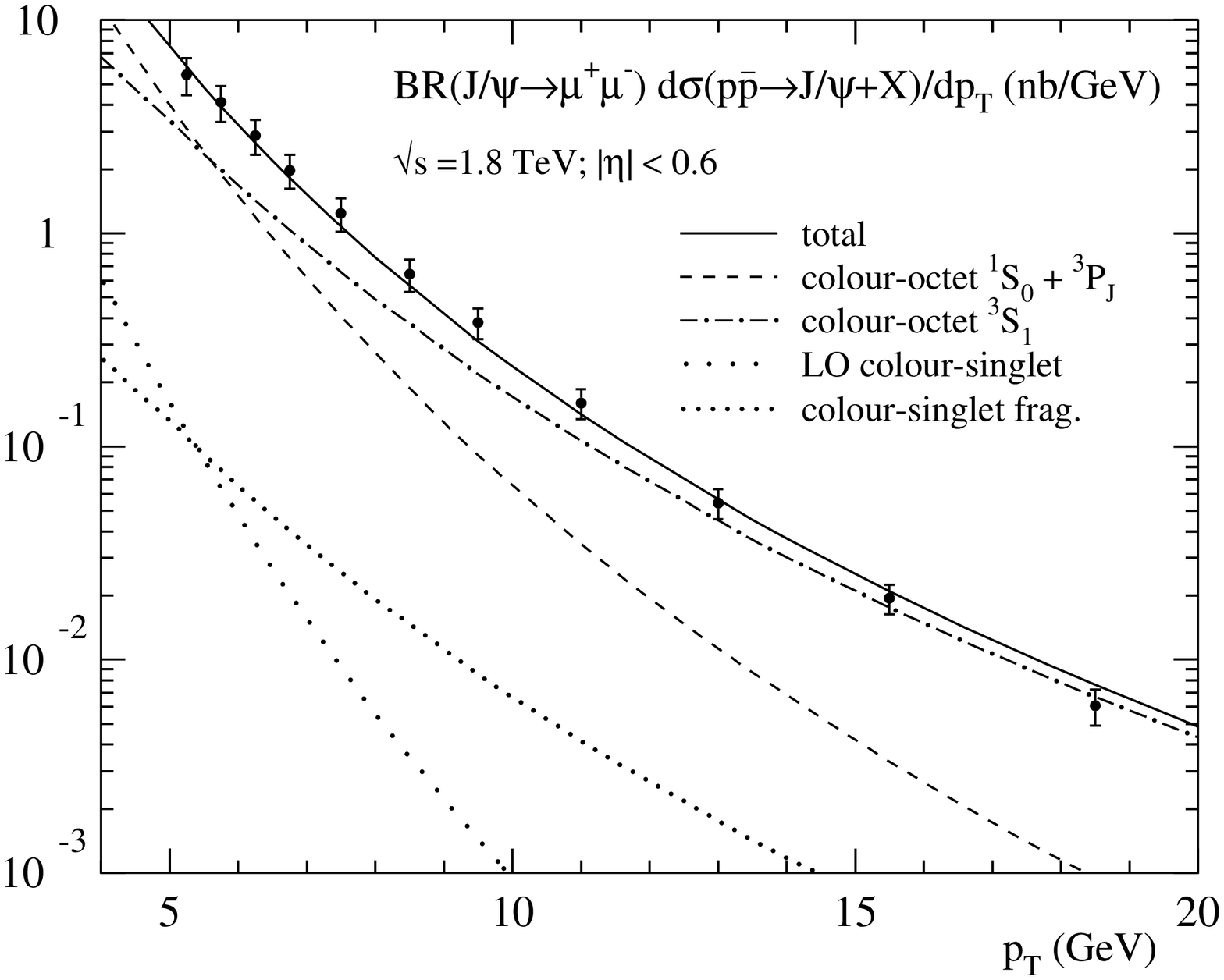}

(a)
\end{center}
\end{minipage}
\hfil
\begin{minipage}[c]{2.7in}
\begin{center}
  \includegraphics[width=1.9in,height=2.6in,angle=270]{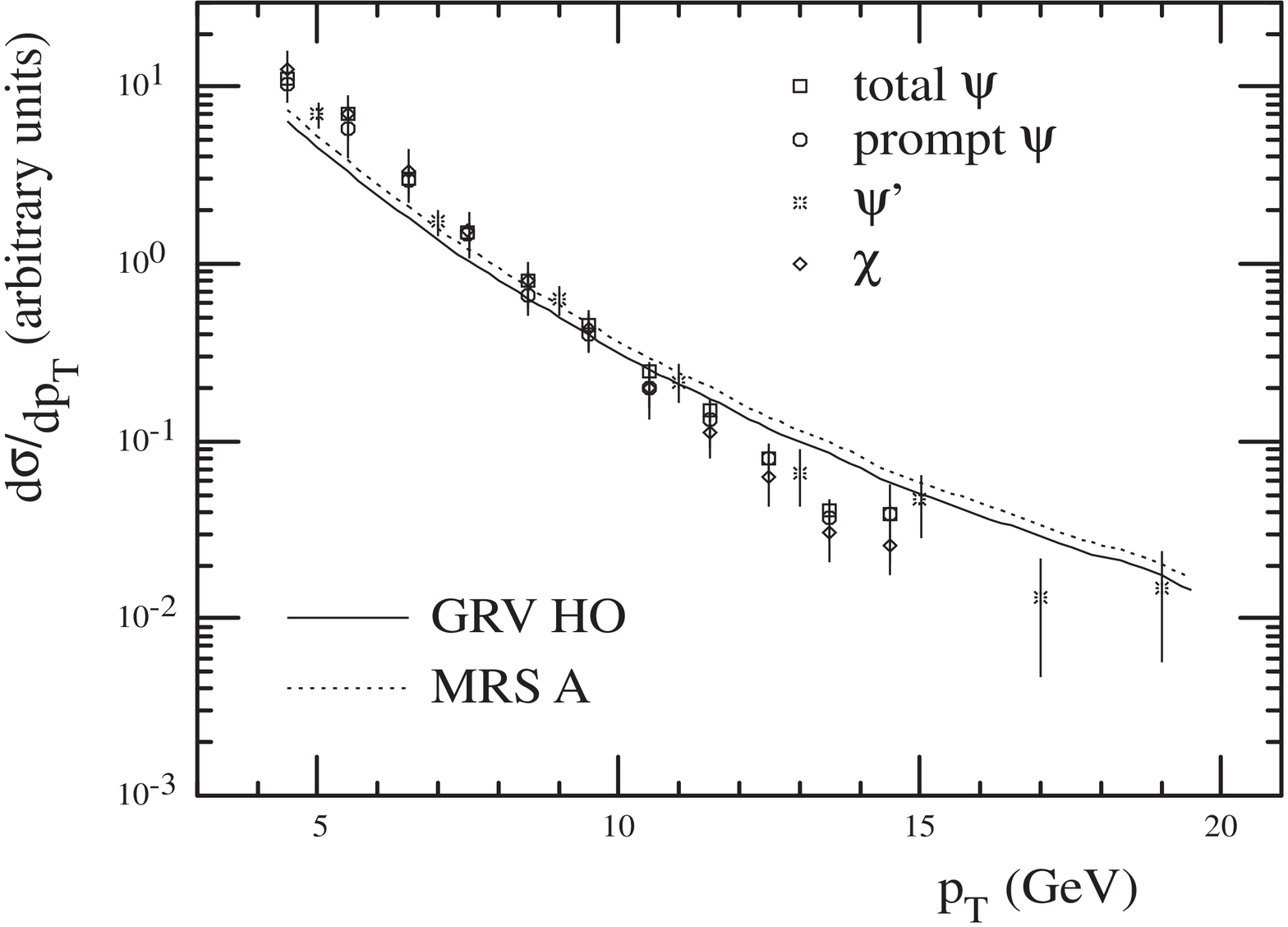}

(b)
\end{center}
\end{minipage}
\caption{Charmonium cross section as a function of $p_T$ along with 
the CDF data points \protect\cite{cdf-jpsi-run1} and 
the theory curves from 
NRQCD model from Ref.~\protect\cite{nrqcd-review} (a), and 
CEM from Ref.~\protect\cite{cem2} (b).}
\label{fig1}
\end{figure}

\vskip 0.1in
\noindent{\it Color-singlet model}\ \
The color-singlet model (CSM) assumes that only a color singlet 
heavy quark pair with the right quantum number can become a 
quarkonium of the same quantum number and the transition 
from the pair to a meson is given 
by the quarkonium wave function \cite{csm}. 
By neglecting the dependence on the pair's relative momentum 
in $\sigma_{A+B\rightarrow Q\bar{Q}[n]+X}$, 
$\int d\Gamma_{Q\bar{Q}}\, F_{Q\bar{Q}\to H}$ in Eq.~(\ref{qq-fac})
is set to equal the matrix element in Eq.~(\ref{F-nonlocal}) 
evaluated at $y_i (\tilde{y}_i)=0$, which is proportional to the 
square of coordinate-space quarkonium wave function
at the origin, $|R_H(0)|^2$ \cite{qwg-review},
and therefore,
\begin{equation}
\sigma^{\rm CSM}_{A+B\rightarrow H+X} 
\propto
\sigma_{A+B\rightarrow Q\bar{Q}[H]+X}(m_Q)\,
|R_H(0)|^2\, .
\end{equation}
The same wave function appears in both production and decay, and 
the model provides absolutely normalized predictions.  It works well 
for J/$\psi$ production in deep inelastic scattering, photon 
production, and some low energy experiments \cite{qwg-review}, 
but fails to predict the CDF data, 
as demonstrated by the dotted lines in 
Fig.~\ref{fig1}(a) \cite{nrqcd-review}.

\vskip 0.1in
\noindent{\it Color evaporation model}\ \
The color evaporation model (CEM) takes a very different approach 
to handle the nonperturbative transition.  It assumes
that all $Q\bar{Q}$ pairs with invariant mass less than the threshold of 
producing a pair of open-flavor heavy mesons, 
regardless their color, spin, and invariant mass, 
have the same probability to become a quarkonium \cite{cem}.  
That is, the $F_{Q\bar{Q}[n]\to H}$ in Eq.~(\ref{qq-fac}) is 
a constant for a given quarkonium state, $H$, and therefore,
\begin{equation}
\sigma^{\rm CEM}_{A+B\rightarrow H+X} 
\approx 
f_H\, 
\int_{2m_Q}^{2M_Q} dm_{Q\bar{Q}}\
\sigma_{A+B\rightarrow Q\bar{Q}+X}(m_{Q\bar{Q}})
\label{cem-fac}
\end{equation}
with open-flavor heavy meson mass $M_Q$ and a constant $f_H$ \cite{cem}. 
With a proper choice of $m_Q$ and $f_H$, 
the model gives a reasonable description of almost all data 
including the CDF data, as shown in Fig.~\ref{fig1}(b).

\begin{figure}
\begin{minipage}[c]{2.7in}
\begin{center}
  \includegraphics[width=2.6in,height=1.8in]{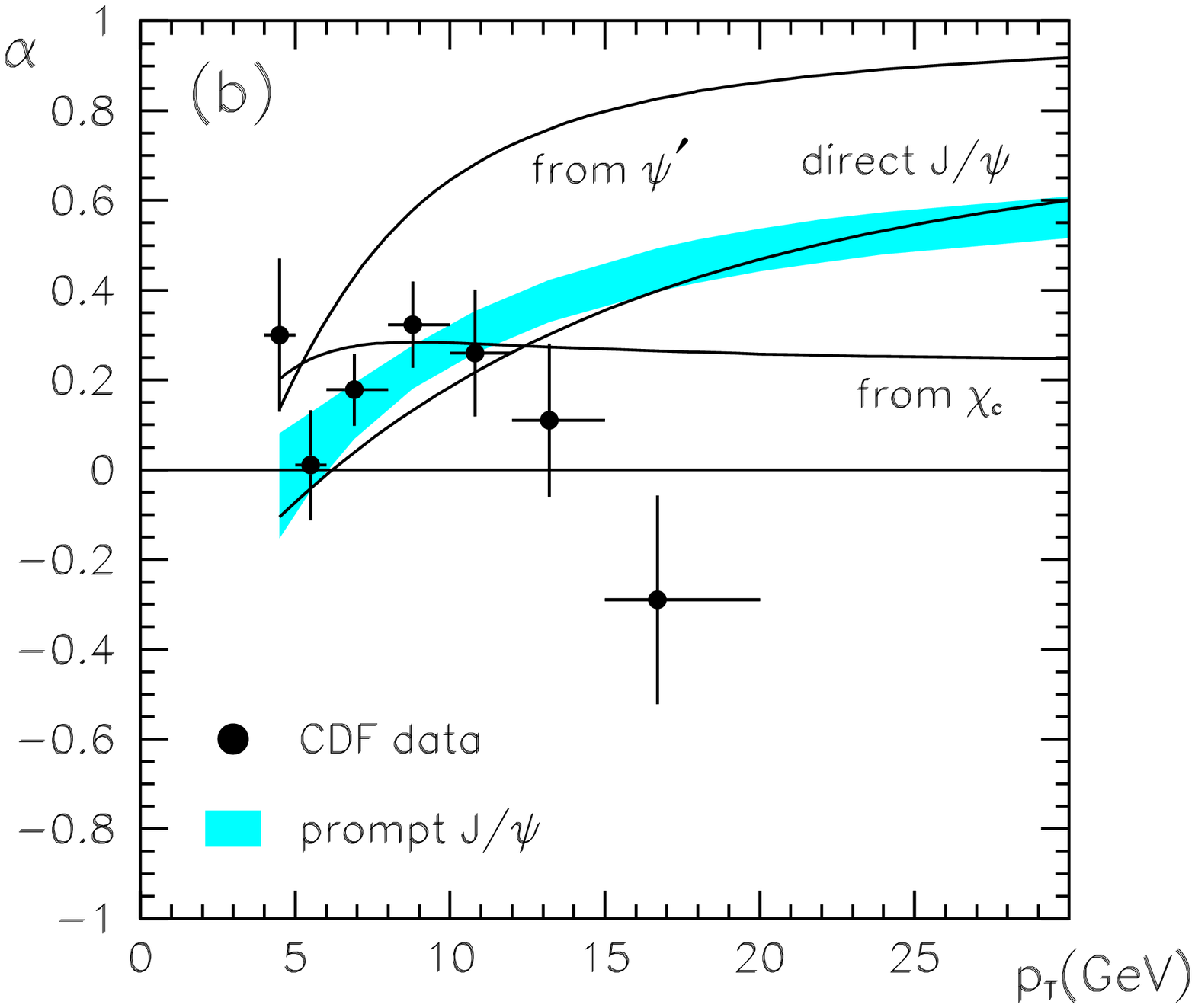}

(a)
\end{center}
\end{minipage}
\hfil
\begin{minipage}[c]{2.7in}
\begin{center}
  \includegraphics[width=2.6in,height=1.8in]{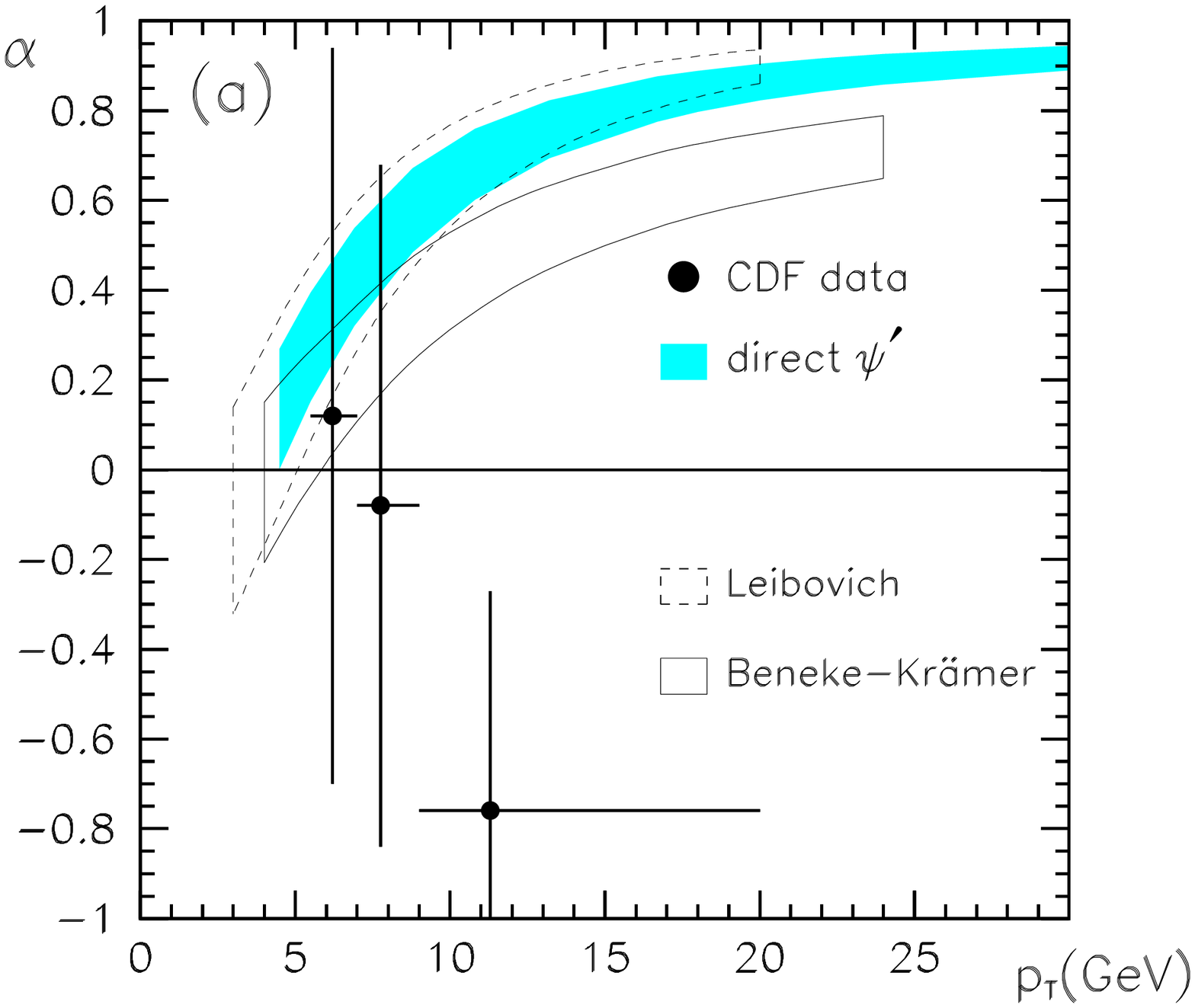}

(b)
\end{center}
\end{minipage}
\caption{From Ref.~\protect\cite{nrqcd-review}, NRQCD 
predictions of charmonium polarizations are compared with  
the CDF data \protect\cite{cdf-pol}.}
\label{fig2}
\end{figure}

\vskip 0.1in
\noindent{\it Nonrelativistic QCD model}\ \
The Nonrelativistic QCD (NRQCD) model is based on the fact that
the typical heavy quark rest-frame kinetic energy and binding energy,
$m_Q v^2$, in a heavy quarkonium is much smaller than the heavy quark 
mass.  The model separates the physics at scales of order $m_Q$ and 
higher from the dynamics of the binding by using NRQCD, 
an effective field theory \cite{bbl-nrqcd}.  
It provides a systematic prescription to calculate the physics
at $m_Q$ order-by-order in powers of $\alpha_s$, and
expands the nonperturbative dynamics in terms of local matrix 
elements ordered in power series of heavy quark velocity $v$ 
\cite{nrqcd-review,bbl-nrqcd},
\begin{equation}
\sigma^{\rm NRQCD}_{A+B\to H+X} 
= \sum_n 
\hat\sigma_{A+B\to Q\bar{Q}[n]+X}\
\langle {\mathcal  O}^H_n\rangle\, ,
\label{nrqcd-fac}
\end{equation}
where the ${\mathcal O}^{H}_n$ are NRQCD operators
for the state $H$ \cite{bbl-nrqcd}, 
\begin{equation}
{\mathcal O}^H_n(0)
=
\chi^\dagger(0)\, {\mathcal K}_n\, \psi(0)\, 
\left(a^\dagger_Ha_H\right)\,
\psi^\dagger(0)\, {\mathcal K}'_n\, \chi(0)\, ,
\label{On-def}
\end{equation}
where $a^\dagger_H$ is the creation operator for $H$, 
$\chi$ ($\psi$) are two component Dirac spinors, and 
${\mathcal K}_n$ and ${\mathcal K}'_n$ 
are defined in Eq.~(\ref{F-nonlocal}) and  
can also involve covariant derivatives.  
At higher orders in $v$, the operator ${\mathcal O}^{H}_n$ 
can have additional dependence on field strength as well as more
fermion fields.  The factorization in Eq.~(\ref{nrqcd-fac}) could 
be understood from Eq.~(\ref{qq-fac}) by expanding the
$\sigma_{A+B\rightarrow Q\bar{Q}[n]+X}$ at heavy
quark relative momentum, $q=(p_Q-p_{\bar{Q}})/2=0$. The
moments, $\int d\Gamma_{Q\bar{Q}}\ q^N\, F_{Q\bar{Q}\to H}$,
lead to local matrix elements with high powers of $v$. 

The NRQCD model allows every $Q\bar{Q}[n]$ state 
to become a bound quarkonium, 
while the probability is determined by corresponding
nonperturbative matrix elements $\langle {\mathcal  O}^H_n\rangle$.
It in principle includes the physics of CSM.  
Its octet contribution is the most important one for high $p_T$
quarkonium production at collider energies \cite{nrqcd-review}. 
The NRQCD model has been most successful in 
interpreting data 
\cite{qwg-review,nrqcd-review}, as shown in  Fig.~\ref{fig1}(a).

\vskip 0.05in
\noindent{\it Quarkonium polarization and other models}\ \
The key difference between the NRQCD model and the CEM is the 
prediction on quarkonium polarization.  Once the matrix elements
$\langle {\mathcal  O}^H_n\rangle$ are determined, the NRQCD
model can systematically calculate  the polarization
of produced heavy quarkonia. On the other hand, CEM does not
provide a systematic prescription to calculate the polarization
because of the uncontrolled radiations from the quark pair. 
The polarization of quarkonia at large $p_T$ 
was considered a definite test of the NRQCD model 
\cite{nrqcd-review}.  But, the model has failed the test, 
as seen in Fig.~\ref{fig2}, 
if the CDF data hold \cite{cdf-pol}.  
Several improved and new models have been proposed to 
address the issues of polarization \cite{Lansberg:2006dh}.

\begin{figure}
\begin{minipage}[c]{2.4in}
\begin{center}
  \includegraphics[width=1.7in,height=1.4in]{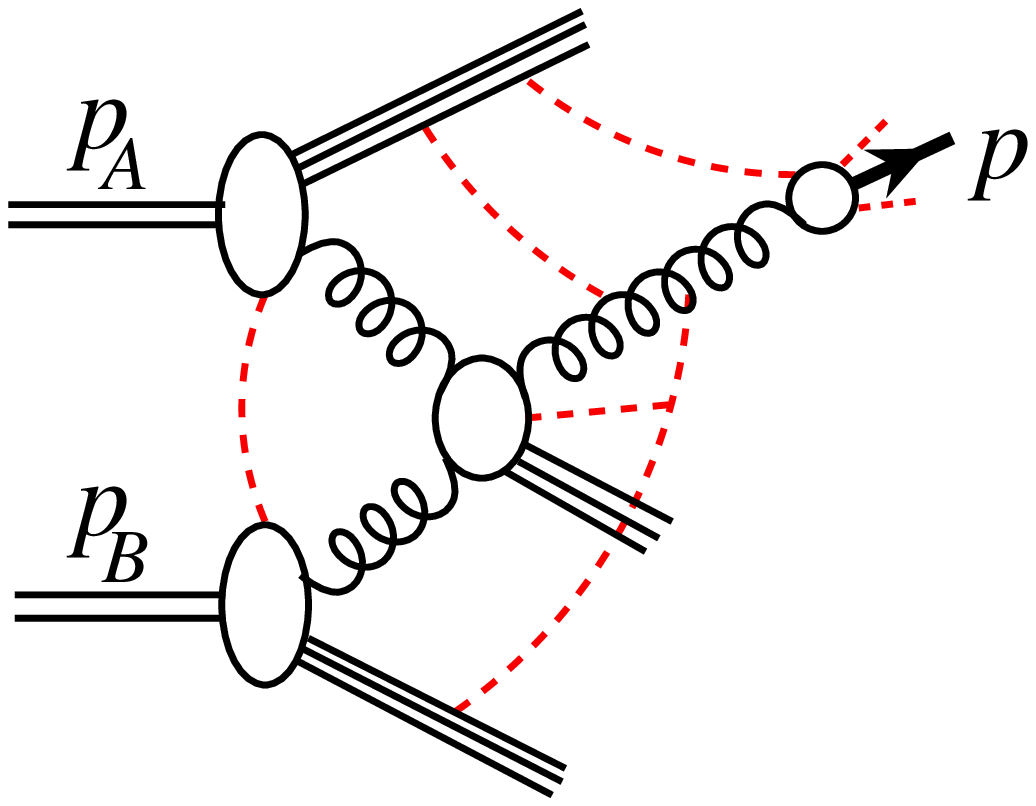}

(a)
\end{center}
\end{minipage}
\hskip 0.2in
\begin{minipage}[c]{2.8in}
\begin{center}
\hskip0.2in
\begin{minipage}[c]{1.0in}
  \includegraphics[width=1.0in,height=0.7in]{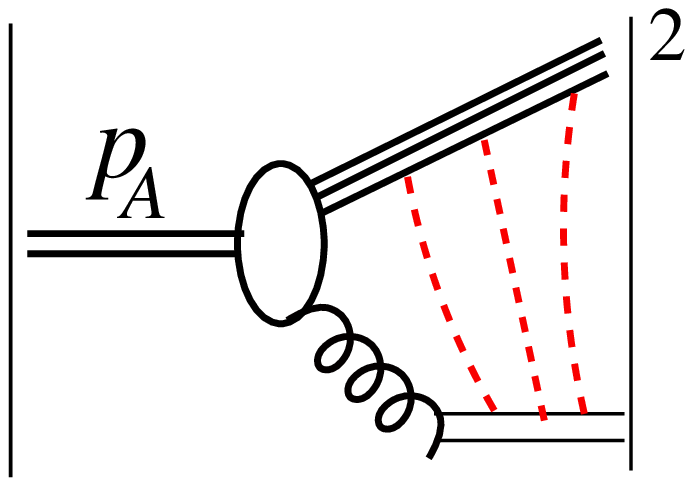}
\end{minipage}
   \hskip 0.1in
   {
    $\otimes $}
   \hskip 0.1in
\begin{minipage}[c]{1.0in}
  \includegraphics[width=1.0in,height=0.7in]{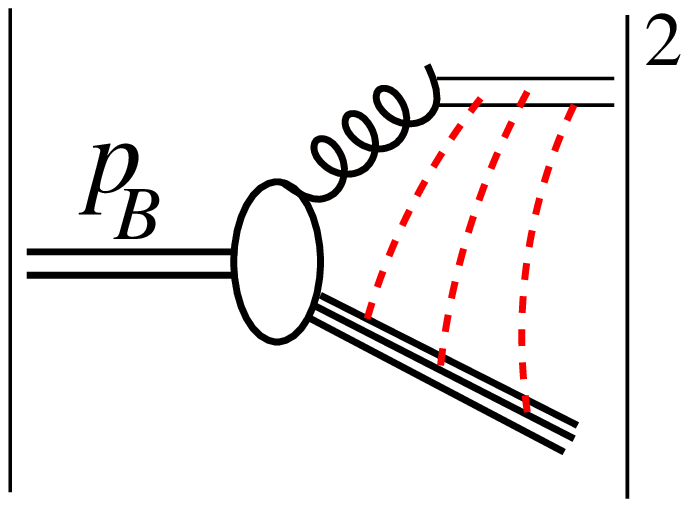}
\end{minipage}

\vskip 0.1in

   {
    $\otimes $}
   \hskip 0.1in
\begin{minipage}[c]{0.8in}
  \includegraphics[width=0.8in,height=0.6in]{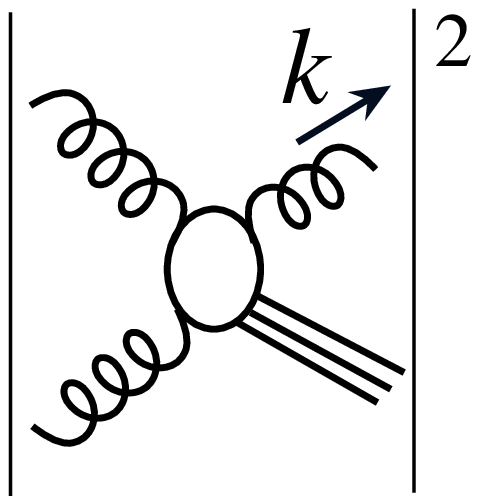}
\end{minipage}
   \hskip 0.1in
   {
    $\otimes $}
   \hskip 0.1in
\begin{minipage}[c]{0.8in}
  \includegraphics[width=0.8in,height=0.6in]{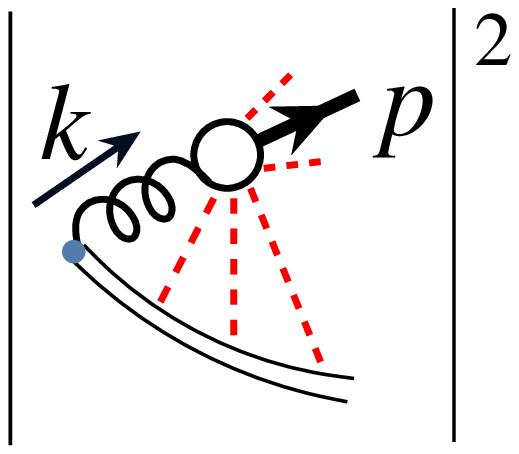}
\end{minipage}

\hskip 0.2in (b)  
\end{center}
\end{minipage}
\caption{Sketch for process $A+B\to H(p)+X$: 
(a) sample scattering amplitude with 
    possible factorization breaking soft interactions 
    indicated by dashed lines, and 
(b) factorization with Wilson lines 
    indicated by thin double lines.}
\label{fig3}
\end{figure}

\section{Factorization}
\label{factorization}

The heavy quarkonium production in hadronic collisions involves both 
perturbative and nonperturbative scales.  The nonperturbative physics 
appears not only in the transition from the heavy quark pair to 
a bound state but also in incoming hadron wave functions.
A typical scattering amplitude for quarkonium production, 
as sketched in Fig.~\ref{fig3}(a), can have soft and 
nonperturbative interactions between incoming hadrons 
as well as between the spectators and the formation process.
These soft interactions may introduce 
process dependence to the nonperturbative matrix elements, and
consequently, spoil the predictive powers of
Eqs.~(\ref{cem-fac}) and (\ref{nrqcd-fac}). 
 
A proof of the factorization needs to: 1) show that the square of the 
scattering amplitude in Fig.~\ref{fig3}(a), after summing over 
all amplitudes with the same initial and final states, 
can be expressed as a convolution of the probabilities, 
as sketched in Fig.~\ref{fig3}(b); each probability represents 
a square of sub-amplitudes and is evaluated at its own
momentum scale(s); 2) show that the piece evaluated at 
perturbative scale(s) is infrared safe and those evaluated 
at nonperturbative scales are universal.  

As argued in Ref.~\cite{heavyquark}, cross sections for producing
on-shell heavy quark pairs can be computed in terms of QCD 
factorization.  Therefore, the right-hand-side of the 
CEM formalism in Eq.~(\ref{cem-fac}) can be factorized into a 
convolution of a calculable partonic cross section of producing 
the heavy quark pair and two universal parton distributions 
from respective hadrons.  However, the factorization
statement here does not provide justification that the 
$F_{Q\bar{Q}[n]\to H}$ in Eq.~(\ref{qq-fac}) 
is independent of the pair's invariant mass 
$m_{Q\bar{Q}}$, spin, and other quantum numbers.  
For quarkonium production at a large 
$p_T$, the $f_H$ should be universal {\it within the model}\ because
soft interactions in Fig.~\ref{fig3}(a)
are suppressed by powers of $1/p_T$.  
When $p_T \ll m_Q$, the $f_H$ may not be universal.

Fully convincing arguments
have not yet been given for NRQCD factorization formalism in
Eq.~(\ref{nrqcd-fac}) \cite{nrqcd-review,bbl-nrqcd}.  
Since the interaction between the beam jet 
and the jet of heavy quark pair should be suppressed by powers of 
$1/p_T$, one might expect the NRQCD factorization formalism to work 
at large $p_T$.  
When $p_T\gg m_Q$, the heavy quarkonium production is similar to the 
single light hadron production, and is dominated by parton 
fragmentation.  The cross section is 
proportional to the universal parton-to-hadron fragmentation 
functions \cite{nqs},
\begin{equation}
\sigma_{A+B\to H+X}(p_T) = 
\sum_i\; \hat\sigma_{A+B\to i+X}(p_T/z,\mu) \otimes
D_{H/i}(z,m_Q,\mu) + {\mathcal O}(m_H^2/p_T^2)\, .
\label{cofact}
\end{equation}
Here, $\otimes$ represents a convolution in the momentum fraction $z$.
The cross section $\hat\sigma_{A+B\to i+X}$ includes all 
information on the incoming state, including convolutions with 
parton distributions of hadrons $A$ and $B$ 
at factorization scale $\mu$, as sketched in 
Fig.~\ref{fig3}(b).  As a necessary condition for
NRQCD factorization in Eq.~(\ref{nrqcd-fac}), 
the following factorization relation, 
\begin{equation}
D_{H/i}(z,m_Q,\mu) = \sum_n\; d_{i\to Q\bar{Q}[n]}(z,\mu,m_Q) \, 
\langle {\mathcal O}^H_n\rangle\, ,
\label{frag-fac}
\end{equation}
is required to be valid to all orders in $\alpha_s$ and 
all powers in $v$-expansion
for all parton-to-quarkonium fragmentation functions \cite{nqs}.
In Eq.~(\ref{frag-fac}), $d_{i\to Q\bar{Q}[n]}$ 
describes the evolution of an off-shell parton into a heavy quark 
pair in state $[n]$, including logarithms of $\mu/m_Q$, and 
should be infrared safe \cite{nqs}.
   
\begin{figure}
\begin{minipage}[c]{1.7in}
\begin{center}
  \includegraphics[width=1.5in,height=1.2in]{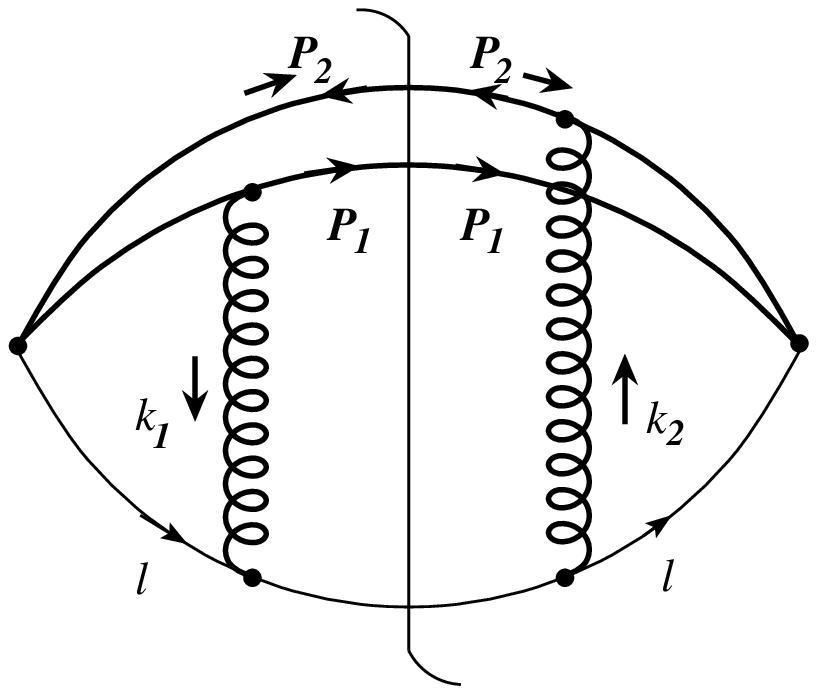}

(I)
\end{center}
\end{minipage}
  \hfil
\begin{minipage}[c]{1.7in}
\begin{center}
  \includegraphics[width=1.5in,height=1.2in]{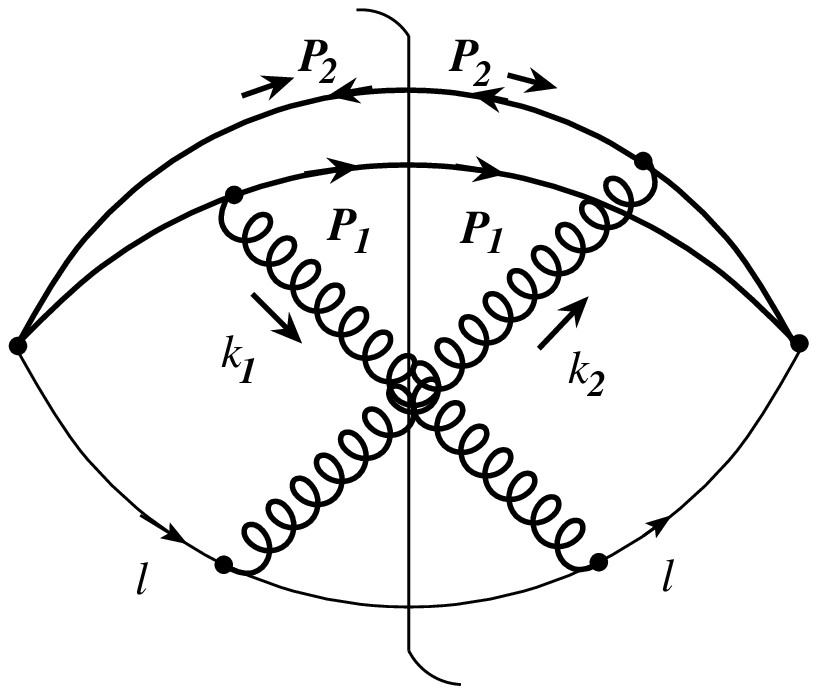}

(II)
\end{center}
\end{minipage}
  \hfil
\begin{minipage}[c]{1.7in}
\begin{center}
  \includegraphics[width=1.5in,height=1.2in]{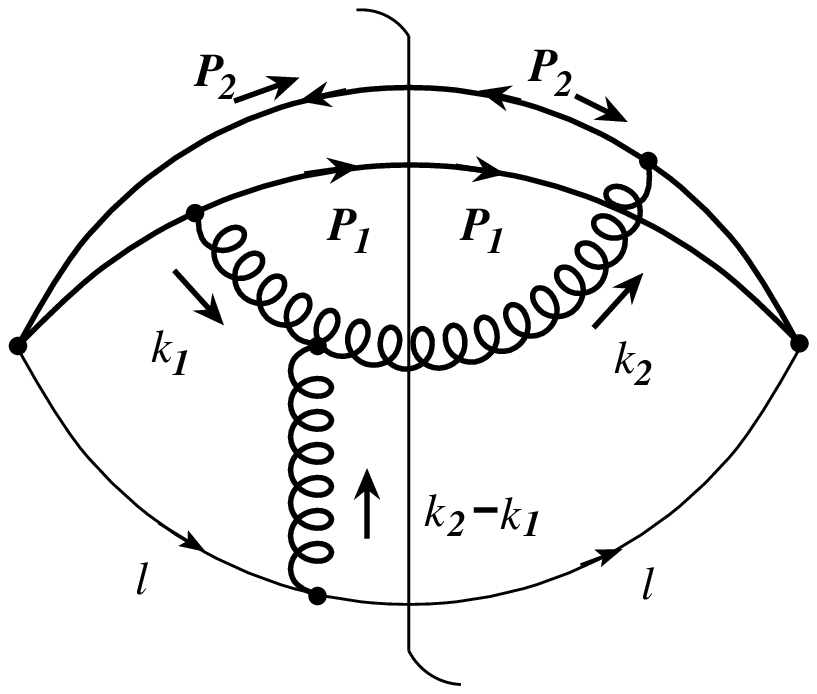}

(III)
\end{center}
\end{minipage}
\caption{Representative NNLO contributions to $g\to Q\bar{Q}$ 
fragmentation in eikonal approximation, see Ref.~\protect\cite{nqs}
for the details.}
\label{fig4}
\end{figure}

The factorization relation in Eq.~(\ref{frag-fac}) 
was tested up to next-to-next-to-leading order (NNLO) in
$\alpha_s$ at $v^2$ order in Ref.~\cite{nqs}, as well as 
at finite $v$ in Ref.~\cite{nqs2}.  
Consider representative NNLO contributions to 
the fragmentation process of transforming a color octet heavy
quark pair to a singlet, as sketched in Fig.~\ref{fig4}.
The individual classes of diagrams in Fig.~\ref{fig4}(I) and (II), 
for which two gluons are exchanged between the
quarks and the Wilson line, satisfy the infrared
cancellation conjecture of Ref.~\cite{bbl-nrqcd}, 
by summing over the possible cuts and connections to quark and
antiquark lines, as do diagrams that have three gluon-eikonal
vertices on the quark pair and one on the  Wilson line \cite{nqs}.  
For Fig.~\ref{fig4}(III) type of diagrams, however, 
with a three-gluon interaction, this cancellation
fails.  By summing over all contributions to second order 
in the relative momentum $q$, it was found that  
the fragmentation function has a noncanceling real pole \cite{nqs}
\begin{equation}
{\mathcal I}^{(8\to 1)}_2(v) 
= - \alpha_s^2\, \frac{1}{3\varepsilon}\, v^2 \, .
\label{v2}
\end{equation}
From Fig.~\ref{fig4}(III), 
this infrared divergence
is not topologically factorizable and 
can not be absorbed into the matrix elements 
$\langle {\mathcal  O}^H_n\rangle$.   
As demonstrated in Ref.~\cite{nqs}, 
as a necessary condition for restoring the NRQCD factorization
at NNLO, the conventional ${\mathcal O}(v^2)$
octet NRQCD production matrix elements
$\langle {\mathcal  O}^H_n\rangle$
must be modified by incorporating Wilson lines 
that make them manifestly gauge invariant,
so that the infrared divergence in Eq.~(\ref{v2}) can be
absorbed into the gauge-completed matrix elements.

For quarkonium production, all heavy quark pairs with invariant mass 
less than a pair of open-flavor heavy mesons could become
a bound quarkonium.  Therefore, the effective velocity of heavy
quark pair in the production could be much larger than that in decay.
For charmonium production, 
charm quark velocity, $v_c\sim |\vec{q}_c|/m_c \leq
\sqrt{(4M_D^2-4m_c^2)/(4m_c^2)} \sim 0.88$, is not small, and 
therefore, the velocity expansion for charmonium production 
may not be a good approximation, unless one can identify and resum
large contributions to all order in $v$ or have a factorized 
formalism at finite $v$.  It was found in Ref.~\cite{nqs2} that
with the gauge-completed matrix elements, 
infrared singularities in the fragmentation function for a color
octet pair to a singlet at NNLO are consistent with NRQCD 
factorization to all orders in $v$ or to a finite $v$,
\begin{equation}
{\mathcal I}^{(8\to 1)}(v)
= \frac{\alpha_s^2}{4\varepsilon} 
\left[ 1-\frac{1}{2f(v)} 
         \ln\left[\frac{1+f(v)}{1-f(v)}\right]
\right] \, ,
\label{allorder}
\end{equation}
where $f(v)=2v/(1+v^2)$.
The result in finite $v$ is remarkably compact and intriguing,
and should encourage further work on the factorization 
theorem.

\section{Summary and outlook}
\label{summary}

After more than 30 years since the discovery of J/$\psi$, we still
have not been able to fully understand the production mechanism of 
heavy quarkonium in high energy collisions, in particular, the 
transition from the heavy quark pair to a bound quarkonium.  
Although there are good reasons for each production model, 
none of the factorized production formalism, including that of the 
NRQCD model, has been proved theoretically. Further work on the
factorization theorem is critical.

The heavy quarkonium production has a ``long'' transition time from 
the produced pairs to bound mesons, and can be a good probe
for the properties of newly formed hot and dense matter of quarks 
and gluons, or the quark-gluon plasma, in ultrarelativistic 
heavy ion collisions \cite{rhic-3yrs,MS-jpsi}, if we can 
calibrate the production.  Since the medium interaction is
sensitive to the quantum state of the proporgating heavy 
quark pair, nuclear matter could be an effective filter 
to distinguish the production mechanism  \cite{qvz-jpsi}.  
Detailed studies of nuclear dependence of heavy quarkonium 
production in hadron-nucleus collisions 
should provide  invaluable information on the 
formation of heavy quarkonia in hadronic collisions.


\begin{theacknowledgments}
The author thanks G.~T.~Bodwin, G.~Nayak, and G.~Sterman 
for discussions, and nuclear theory group at Brookhaven National 
Laboratory for its hospitality during the writing of this contribution.
This work is supported in part by the U.S. Department of Energy, 
Grant No. DE-FG02-87ER40371, and Contract No. DE-AC02-98CH10886.
\end{theacknowledgments}


\end{document}